\documentclass[aps, prl, 
reprint, superscriptaddress, nobibnotes, 
amsmath, amssymb, 
]{revtex4-2}

\usepackage{hyperref}
    \hypersetup{
        colorlinks=true, 
        linkcolor=blue,
        urlcolor=blue,
        citecolor=blue,
        bookmarksopen=true,
    }
\usepackage[all]{hypcap}
\usepackage{graphicx}
\usepackage{latexsym}
\usepackage{bm}
\usepackage{siunitx}
\usepackage[normalem]{ulem}

\usepackage[dvipsnames]{xcolor}
\definecolor{lightgreen}{rgb}{0, 0.7, 0}
\newif\ifSwitch
\newif\ifRed
\newif\ifGreen
\newif\ifHide
\Redtrue
\Greentrue
\newcommand{\Red}[1]{{\ifRed\textcolor{red}{#1}\else\textcolor{black}{#1}\fi}}
\newcommand{\Green}[1]{{\ifGreen\textcolor{lightgreen}{#1}\else\textcolor{black}{#1}\fi}}
\newcommand{\Strike}[1]{\ifHide\relax\ignorespaces\else\sout{#1}\fi}
\Switchtrue
\Redfalse
\Greenfalse
\Hidetrue

\newcommand*{\PLCCO}{${\text{Pr}_{\text{1.3-}x}\text{La}_\text{0.7}\text{Ce}_x\text{CuO}_\text{4-$\delta$}}$}
\newcommand*{\refer}[2]{\hyperref[#1]{\ref{#1}(#2)}}
\newcommand*{\Tc}{$T_\text{c}$}
\newcommand*{\Tp}{T$'$}
\newcommand*{\nFS}{$n_\text{FS}$}
\newcommand*{\EF}{$E_\text{F}$}
\newcommand*{\wave}{$\sim$}
\newcommand*{\Neel}{N$\acute{\text{e}}$el}
\newcommand*{\etal}{\textit{et~al.}}

\begin{document}

\title{Extended superconducting dome of electron-doped cuprates after protect annealing revealed by ARPES}

\author{C.~Lin}
    \email[]{clin@issp.u-tokyo.ac.jp}
    \affiliation{Department of Physics, University of Tokyo, Bunkyo-ku, Tokyo 113-0033, Japan.}
\author{T.~Adachi}
    \affiliation{Department of Engineering and Applied Sciences, Sophia University, Tokyo 102-8554, Japan.}
\author{M.~Horio}
    \affiliation{Department of Physics, University of Tokyo, Bunkyo-ku, Tokyo 113-0033, Japan.}
    \affiliation{Physik-Institut, Universit\"{a}t Z\"{u}rich, Winterthurerstrasse 190, CH-8057 Z\"{u}rich, Switzerland.}
\author{T.~Ohgi} 
    \affiliation{Department of Applied Physics, Tohoku University, Sendai 980-8579, Japan.}
\author{M.A.~Baqiya}
    \affiliation{Department of Applied Physics, Tohoku University, Sendai 980-8579, Japan.}
    \affiliation{Department of Physics, Faculty of Science and Data Analytics, Institut Teknologi Sepuluh Nopember, Surabaya 60111, Indonesia.}
\author{T.~Kawamata}
\author{H.~Sato}
    \affiliation{Department of Applied Physics, Tohoku University, Sendai 980-8579, Japan.}
\author{T.~Sumura}
    \affiliation{Department of Engineering and Applied Sciences, Sophia University, Tokyo 102-8554, Japan.}
\author{K.~Koshiishi}
\author{S.~Nakata}
\author{G.~Shibata}
\author{K.~Hagiwara}
\author{M.~Suzuki}
    \affiliation{Department of Physics, University of Tokyo, Bunkyo-ku, Tokyo 113-0033, Japan.}
\author{K.~Ono}
\author{K.~Horiba}
\author{H.~Kumigashira}
    \affiliation{KEK, Photon Factory, Tsukuba 305-0801, Japan.}
\author{S.~Ideta}
\author{K.~Tanaka}
    \affiliation{UVSOR Facility, Institute for Molecular Science, Okazaki 444-8585, Japan.}
\author{Y.~Koike}
    \affiliation{Department of Applied Physics, Tohoku University, Sendai 980-8579, Japan.}
\author{A.~Fujimori}
    \affiliation{Department of Physics, University of Tokyo, Bunkyo-ku, Tokyo 113-0033, Japan.}
    \affiliation{Department of Applied Physics, Waseda University, Shinjuku, Tokyo 169-8555, Japan.}


\begin{abstract}
The electron-doped cuprates are usually characterized by a more robust antiferromagnetic phase and a much narrower superconducting (SC) dome than those of the hole-doped counterparts. 
Recently, bulk single crystals of \PLCCO{} (PLCCO) prepared by the protect annealing method have been studied extensively and revealed many intriguing properties that were different from those obtained from samples annealed by the conventional methods.
Here, we report on a systematic angle-resolved photoemission spectroscopy study of PLCCO single crystals after protect annealing.
The results indicate that the actual electron concentration (\nFS{}) estimated from the Fermi-surface area is significantly larger than the Ce concentration $x$ and the new \nFS{}-based SC dome of PLCCO is more extended towards the overdoped side than the $x$-based SC dome derived for samples prepared using the conventional annealing method.
\Red{\Strike{The similarity between the new \nFS{}-based SC dome and that of the hole-doped cuprate ${\text{La}_{\text{2-}x}\text{Sr}_x\text{CuO}_\text{4}}$ further provides a clue for understanding the reported electron-hole symmetry/asymmetry of the cuprate phase diagram.}}
\end{abstract}

\maketitle

\begin{figure}[!ht]
    \includegraphics[width=0.45\textwidth]{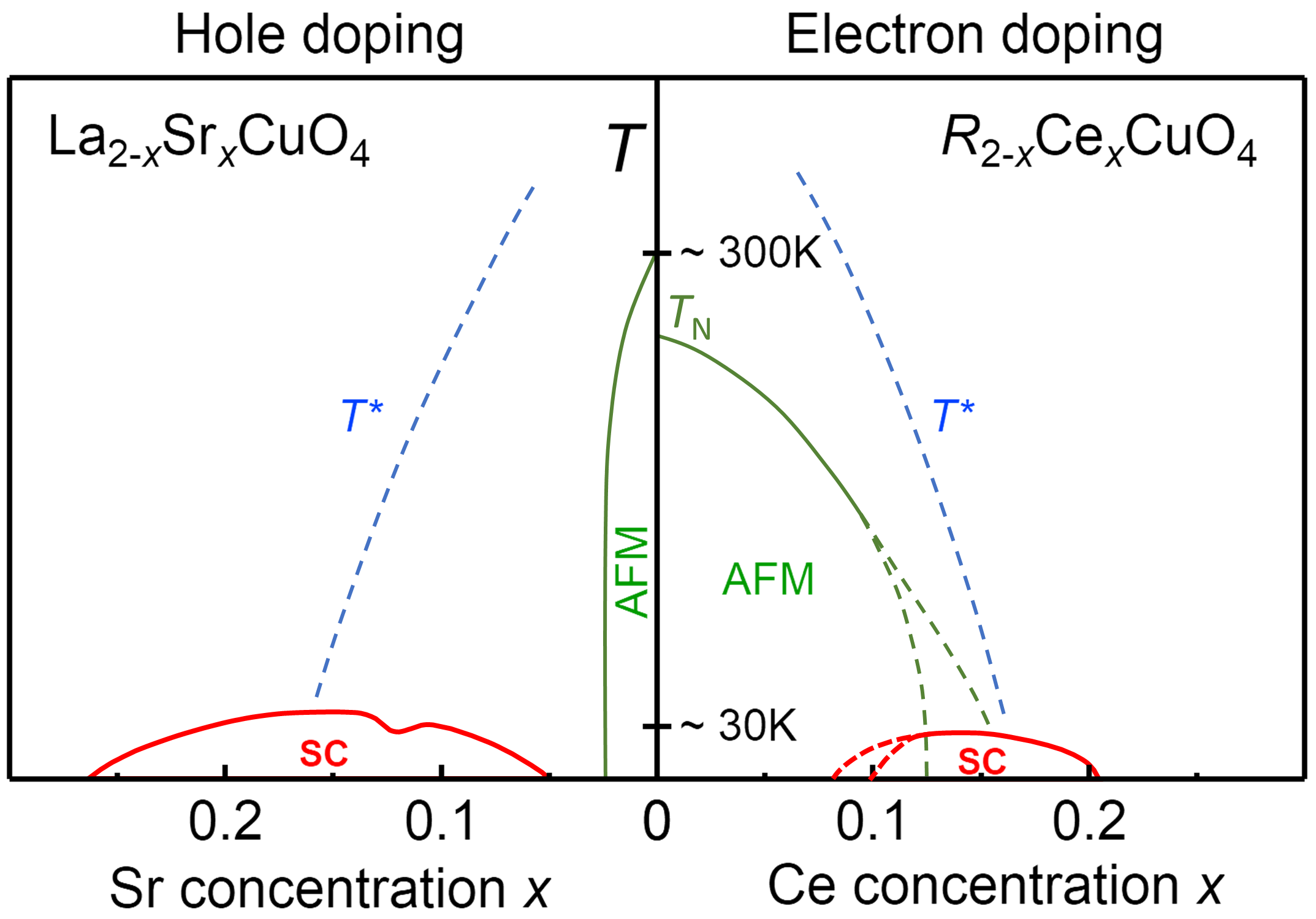}
    \caption[Traditional Cuprate Phase Diagram]{
    Phase diagram of the hole- and electron-doped cuprate superconductors, taking an example of the ${\text{La}_{\text{2-}x}\text{Sr}_x\text{CuO}_\text{4}}$ (LSCO) and ${\textit{R}_{\text{2-}x}\text{Ce}_x\text{CuO}_\text{4}}$ (\textit{R} represents a rare earth element) systems~\cite{Armitage_Rev.Mod.Phys.823_2010}.
    $T_N$ and $T^*$ indicate the \Neel{} temperature and the approximate onset temperature of the pseudogap, respectively.
    The dashed green and red curves \Red{are estimated from the literature}~\cite{Luke_Phys.Rev.B4213_1990,Fujita_Phys.Rev.B671_2003,Fujita_Phys.Rev.Lett.10110_2008,Motoyama_Nature4457124_2007,Uefuji_PhysicaC:Superconductivity357-360_2001, Matsuda_Phys.Rev.B4521_1992, Uefuji_PhysicaC:Superconductivity378-381_2002,Mang_Phys.Rev.Lett.932_2004,Takagi_Phys.Rev.Lett.6210_1989,Luke_Phys.Rev.B4213_1990,Motoyama_Nature4457124_2007,Fujita_Phys.Rev.B671_2003,Fujita_Phys.Rev.Lett.10110_2008,Uefuji_PhysicaC:Superconductivity357-360_2001,Song_Phys.Rev.Lett.11813_2017,Naito_Jpn.J.Appl.Phys.396A_2000,Sawa_Phys.Rev.B661_2002,Krockenberger_Phys.Rev.B(RapidComm.)776_2008} denoting the uncertain ranges of the antiferromagnetic (AFM) phase~\cite{Luke_Phys.Rev.B4213_1990,Fujita_Phys.Rev.B671_2003,Fujita_Phys.Rev.Lett.10110_2008,Motoyama_Nature4457124_2007,Uefuji_PhysicaC:Superconductivity357-360_2001, Matsuda_Phys.Rev.B4521_1992, Uefuji_PhysicaC:Superconductivity378-381_2002,Mang_Phys.Rev.Lett.932_2004} and the superconducting (SC) phase~\cite{Takagi_Phys.Rev.Lett.6210_1989,Luke_Phys.Rev.B4213_1990,Motoyama_Nature4457124_2007,Fujita_Phys.Rev.B671_2003,Fujita_Phys.Rev.Lett.10110_2008,Uefuji_PhysicaC:Superconductivity357-360_2001,Song_Phys.Rev.Lett.11813_2017,Naito_Jpn.J.Appl.Phys.396A_2000,Sawa_Phys.Rev.B661_2002,Krockenberger_Phys.Rev.B(RapidComm.)776_2008} in the electron-doped cuprates, respectively.
    \label{Phase_Diagram}
    }
\end{figure}

The discovery of the electron-doped cuprates guides a new way in unveiling the controversial physics of high-temperature superconductivity and motivates extensive experimental and theoretical studies for decades~\cite{Tokura_Nature3376205_1989,Armitage_Rev.Mod.Phys.823_2010}.
Other than the structure differences, the most dramatic distinction between the hole- and electron-doped cuprates embodies in their temperature-doping phase diagram as shown in Fig.~\ref{Phase_Diagram}.
On the hole doping side, a tiny amount \wave{} 3\% of hole doping suppresses the antiferromagnetic (AFM) insulating phase and the systems become superconducting (SC) at \wave{} 5\% hole doping in most cuprate families~\cite{Damascelli_Rev.Mod.Phys.752_2003,Vishik_PNAS10945_2012,Barisic_PNAS11030_2013,Keimer_Nature5187538_2015}.
On the other hand, on the electron doping side, previous numerous studies revealed that the AFM phase is much more robust~\cite{Luke_Phys.Rev.B4213_1990,Fujita_Phys.Rev.B671_2003,Fujita_Phys.Rev.Lett.10110_2008,Motoyama_Nature4457124_2007} and persists even up to \wave{} 15\% Ce doping~\cite{Uefuji_PhysicaC:Superconductivity357-360_2001, Matsuda_Phys.Rev.B4521_1992, Uefuji_PhysicaC:Superconductivity378-381_2002,Mang_Phys.Rev.Lett.932_2004}.
A superconductivity emerges at a Ce doping level varying from \wave{} 10\% to 14\% in bulk crystals~\cite{Takagi_Phys.Rev.Lett.6210_1989,Luke_Phys.Rev.B4213_1990,Motoyama_Nature4457124_2007,Fujita_Phys.Rev.B671_2003,Fujita_Phys.Rev.Lett.10110_2008,Uefuji_PhysicaC:Superconductivity357-360_2001,Song_Phys.Rev.Lett.11813_2017}, and \wave{} 6\% to 14\% in thin films~\cite{Naito_Jpn.J.Appl.Phys.396A_2000,Sawa_Phys.Rev.B661_2002,Krockenberger_Phys.Rev.B(RapidComm.)776_2008}.
The doping range of the SC dome, which varies from \wave{} 5\% to \wave{} 15\% for electron-doped cuprates~\cite{Takagi_Phys.Rev.Lett.6210_1989,Luke_Phys.Rev.B4213_1990,Motoyama_Nature4457124_2007,Fujita_Phys.Rev.B671_2003,Fujita_Phys.Rev.Lett.10110_2008,Uefuji_PhysicaC:Superconductivity357-360_2001,Naito_Jpn.J.Appl.Phys.396A_2000,Sawa_Phys.Rev.B661_2002,Krockenberger_Phys.Rev.B(RapidComm.)776_2008}, is also considerably more restricted than that of \wave{} 22\% in the hole-doped case~\cite{Damascelli_Rev.Mod.Phys.752_2003,Vishik_PNAS10945_2012,Barisic_PNAS11030_2013,Keimer_Nature5187538_2015}.
\Red{The controversial reports of the phase boundaries in the electron-doped cuprates (see the dashed lines in Fig.~\ref{Phase_Diagram}) may result from the difficulty in synthesizing samples, particularly in controlling the amount of inhomogeneous apical oxygen.
In spite of the controversy,}
these differences of the phase diagram between the hole- and electron-doped cuprates imply that hole doping and electron doping may affect the electronic structure in different manners.

The electron-doped cuprates are characterized by the \Tp{}-type structure, in which the Cu atom is surrounded by four oxygen atoms in the square-planer manner, instead of octahedral manner by six oxygen atoms in the T-type structure of the hole-doped counterparts ${\text{La}_{\text{2-}x}\text{Sr}_x\text{CuO}_\text{4}}$ (LSCO).
Another hallmark of the electron-doped cuprates is the indispensable role of annealing, i.e., as-grown samples are AFM regardless of dopant concentration and superconductivity emerges only after annealing~\cite{Tokura_Nature3376205_1989}.
Nevertheless, the precise effects of the reduction annealing remain unclear~\cite{Armitage_Rev.Mod.Phys.823_2010,Horio_Nat.Commun.7_2016}. 
Historically the most widely acknowledged impact of annealing is the removal of excess and superconductivity-harmful~\cite{Xu_Phys.Rev.B532_1996} apical oxygen atoms in the \Tp{}-type structure~\cite{Armitage_Rev.Mod.Phys.823_2010}, even though the fraction of reduced oxygen is small~\cite{Takayama-Muromachi_PhysicaC:Superconductivity1595_1989,Moran_PhysicaC:Superconductivity1601_1989,Tarascon_Phys.Rev.B407_1989}.
Meanwhile another scenario has also been proposed that annealing may create a secondary phase and repair Cu vacancies which may exist in as-grown samples~\cite{Kurahashi_J.Phys.Soc.Jpn.713_2002,Mang_Phys.Rev.B709_2004,Kang_Nat.Mater.63_2007}.

Since the reduction annealing is crucial for realizing the superconductivity in the electron-doped cuprates, different annealing methods may result in distinct physical properties.
For bulk single crystals, conventional annealing procedures~\cite{Armitage_Rev.Mod.Phys.823_2010} can lead to over-reduction of the surfaces and even decompose the crystals under strong reduction conditions.
Differently from the conventional processes, Adachi \etal{}~\cite{Adachi_J.Phys.Soc.Jpn.826_2013,Adachi_J.Phys.Soc.Jpn.8511_2016,Adachi_Condens.Matter23_2017} recently synthesized bulk SC single crystals of \PLCCO{} (PLCCO, $x$ = 0.05, 0.10, and 0.15) with \Tc{} as high as \wave{} 27 K, by utilizing an improved ``protect annealing'' method~\cite{Brinkmann_Phys.Rev.Lett.7424_1995}.
In the protect annealing procedures, single crystals are protected from the over-reduction by covering them by polycrystalline powders of the same composition, and one can anneal the samples under stronger reduction conditions without being decomposed and the oxygen content becomes more homogeneous, giving rise to a higher \Tc{}.
Inspired by the \Tc{} enhancement of PLCCO single crystals after the improved protect annealing, extensive studies using various techniques have been conducted recently~\cite{Adachi_J.Phys.Soc.Jpn.826_2013,Adachi_J.Phys.Soc.Jpn.8511_2016,Horio_Nat.Commun.7_2016,Ohnishi_J.Phys.Soc.Jpn.874_2018,Horio_ArXiv180104247Cond-Mat_2018,Horio_J.Phys.:Condens.Matter3050_2018,Horio_Phys.Rev.B1005_2019}.
For example, angle-resolved photoemission spectroscopy (ARPES) measurements on protect-annealed PLCCO samples have shown that the signature of the AFM correlations, namely, the ``AFM pseudogap'' is strongly suppressed for $x$ = 0.10~\cite{Horio_Nat.Commun.7_2016}, the Ce content at which previous studies obtained from conventionally annealed electron-doped cuprates all favored the existence of the AFM pseudogap on the Fermi surface created by the band folding~\cite{Matsui_Phys.Rev.Lett.951_2005,Armitage_Phys.Rev.Lett.8714_2001,Matsui_Phys.Rev.B7522_2007,Ikeda_Phys.Rev.B801_2009,Song_Phys.Rev.Lett.11813_2017}.
The striking differences between the conventionally annealed and protect-annealed samples demonstrate again the crucial role of reduction annealing in the electron-doped cuprates.
As a natural consequence of annealing, one would anticipate the addition of electrons resulting from oxygen reduction.
Actually, by improved elaborate annealing methods, superconductivity emerges even at zero Ce doping level in thin films~\cite{Tsukada_SolidStateCommunications1337_2005, Tsukada_PhysicaC:Superconductivity426-431_2005,Matsumoto_PhysicaC:Superconductivity46915_2009,Krockenberger_Sci.Rep.3_2013} and bulk polycrystals~\cite{Takamatsu_Appl.Phys.Express57_2012, Takamatsu_PhysicsProcedia58SupplementC_2014}.
These new annealing methods not only give rise to the realization of superconductivity in the parent compounds without cation substitution~\cite{Wei_Phys.Rev.Lett.11714_2016, Horio_Phys.Rev.Lett.12025_2018, Horio_Phys.Rev.B(RapidComm.)982_2018,Lin_J.Phys.Soc.Jpn.8811_2019}, but may also lead to a total electron concentration that is larger than the Ce doping level~\cite{Kuroshima_PhysicaC:Superconductivity392-396_2003,Mang_Phys.Rev.Lett.932_2004,Horio_Nat.Commun.7_2016,Song_Phys.Rev.Lett.11813_2017,Horio_Phys.Rev.B(RapidComm.)982_2018}.
In the ARPES study focusing on the protect-annealed PLCCO with $x$ = 0.10 and \Tc{} \wave{} 27 K~\cite{Horio_Nat.Commun.7_2016}, the actual electron concentration estimated from the Fermi-surface area (\nFS{})  was found to be as high as 0.18. 
A subsequent ARPES study on conventionally annealed ${\text{Pr}_{\text{1-}x}\text{La}\text{Ce}_x\text{CuO}_\text{4}}$ [PLCCO (La1.0)] with $x$ = 0.10, 0.15, and 0.18~\cite{Song_Phys.Rev.Lett.11813_2017} also indicated that the reduction (oxidization) injects electrons (holes) into the system.
While AFM correlation still exists in the entire doping range, a new \nFS{}-based phase diagram was proposed and shows a dome-like SC region implying the possible absence of asymmetry between the phase diagrams of hole- and electron-doped cuprates.
Comparing these two studies, one can see that, in the protect-annealed PLCCO~\cite{Adachi_J.Phys.Soc.Jpn.826_2013,Adachi_J.Phys.Soc.Jpn.8511_2016,Horio_Nat.Commun.7_2016}, \Tc{} is higher than that of the conventionally annealed samples~\cite{Song_Phys.Rev.Lett.11813_2017}, and no AFM pseudogap was found.
It is also possible that more electrons are doped by utilizing the novel protect annealing method, with less oxygen inhomogeneity not only on surfaces but also in bulk~\cite{Horio_Nat.Commun.7_2016}.
A systematic ARPES study is then needed to elucidate the possible new phase diagram of protect-annealed electron-doped cuprates.

\begin{table*}[!hbt]
    \caption[PLCCO sample compositions and annealing conditions]{
    Sample compositions and annealing conditions.
    \label{PLCCO_Conditions}
    }
    \begin{ruledtabular}
    \begin{tabular}{cccccc}
    {\textit{x}}  &$T_{c}$ {(K)} &{Reduction status}\footnote{UR: under-reduced, OP: optimally-reduced, OR: over-reduced.}
    & \textup{Annealing step}\footnote{PA: protect annealing, LTA: low-temperature annealing, DA: dynamic annealing.} & \textup{Annealing temperature} (\SI{}{\degreeCelsius})& \textup{Annealing time}\footnote{The number after ``$\times$'' is the number of cycles in the dynamic annealing.}\\
    \hline
    0.10 &26 &UR & PA $\rightarrow$ LTA $\rightarrow$ DA & 800 $\rightarrow$ 400 $\rightarrow$ 500 &    \multicolumn{1}{l}{24 h $\rightarrow$ 24 h $\rightarrow$ 4 h $\times$ 6}\\
    0.10 &28 &OP & PA $\rightarrow$ LTA $\rightarrow$ DA & 800 $\rightarrow$ 400 $\rightarrow$ 500 &    \multicolumn{1}{l}{24 h $\rightarrow$ 24 h $\rightarrow$ 4 h $\times$ 12}\\
    0.15 &16-19 &OR & PA $\rightarrow$ LTA $\rightarrow$ DA & 800 $\rightarrow$ 400 $\rightarrow$ 500 &    \multicolumn{1}{l}{24 h $\rightarrow$ 24 h $\rightarrow$ 4 h $\times$ 6}\\
    0.17 &6 &OP & \multicolumn{1}{l}{PA $\rightarrow$ LTA} & \multicolumn{1}{l}{\hspace{18pt}  900 $\rightarrow$ 500} &    \multicolumn{1}{l}{12 h $\rightarrow$ 12 h}\\
    \end{tabular}
    \end{ruledtabular}
\end{table*}

\begin{figure*}[!bt]
    \includegraphics[width=\textwidth]{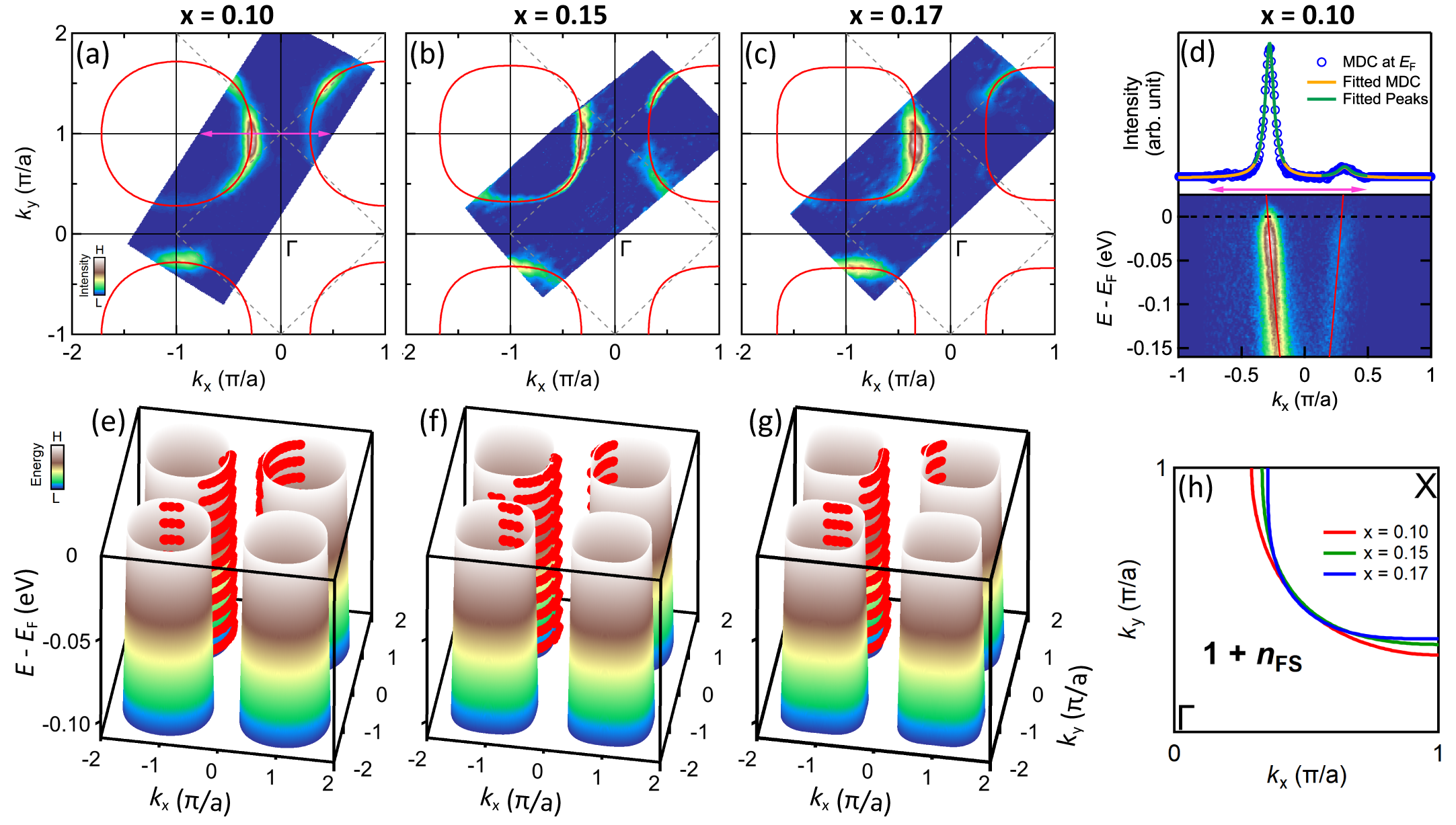}
    \caption[Fermi surfaces and tight-binding fittings for $x$ = 0.10, 0.15, and 0.17]{
    Fermi surfaces and tight-binding fitting of protect-annealed \PLCCO{} (PLCCO) and ${\text{Pr}_{\text{1-}x}\text{La}\text{Ce}_x\text{CuO}_\text{4}}$ [PLCCO (La1.0)]. 
    (a) Optimally-reduced PLCCO with $x$ = 0.10 and \Tc{} = 28 K.
    (b) \Green{\Strike{Over-reduced}} PLCCO with $x$ = 0.15 and \Tc{} = 19 K.
    (c) PLCCO (La1.0) with $x$ = 0.17 and \Tc{} = 6 K.
    The AFM Brillouin-zone boundaries are indicated by the dashed lines.
    The Fermi-surface intensity is integrated over \EF{} $\pm 10$ meV.
    (d) \Red{A representative dispersion and momentum distribution curve (MDC) at \EF{} along the magenta arrow indicated in (a) for PLCCO with $x$ = 0.10.
    The fitted peaks and resulting fitted MDC are also overlapped for comparison.
    See the text for the fitting procedure.
    (e) - (g),
    3D plot of the band dispersions including different energies (red markers) and the corresponding tight-binding fitting (false color plot) for $x$ = 0.10, 0.15, and 0.17, respectively.
    The resulting tight-binding Fermi surfaces are shown in (a) - (c) as red curves.
    }
    (h) Summarized tight-binding fitted Fermi surfaces in (a) - (c).
    The Fermi-surface region of which the area is used to estimate the actual doped electron concentration is indicated by 1 + \nFS{}.
    Details of the sample compositions and annealing conditions are summarized in Table~\ref{PLCCO_Conditions}.
    \Red{Fitted tight-binding parameters are summarized in Table~\ref{TB_Parameters}.}
    \label{Fermi_Surface}
    }
\end{figure*}

In this Letter, we report a systematic ARPES study of PLCCO single crystals after protect annealing [$x$ = 0.10, 0.15, and 0.17 (La1.0)].
The results indicate that the actual doped electron concentration \nFS{} is larger than the Ce doping level $x$ by \wave{} 0.08 $e$/Cu.
The improved annealing method dopes the system with more electrons and the new \nFS{}-based SC dome of PLCCO is more extended towards the overdoped side than that of the conventionally annealed samples~\cite{Song_Phys.Rev.Lett.11813_2017}.
The similarity between the new \nFS{}-based SC dome and that of hole-doped LSCO has thus become clearer and provides a clue for understanding the symmetry/asymmetry of the cuprate phase diagram.

High-quality single crystals of \PLCCO{} ($x$ = 0.10, 0.15) with \Tc{} varying from 16 K to 28 K and ${\text{Pr}_{\text{1.0-}x}\text{La}_\text{1.0}\text{Ce}_x\text{CuO}_\text{4}}$ ($x$ = 0.17) with \Tc{} of 6 K were synthesized by the traveling-solvent floating-zone method~\cite{Adachi_J.Phys.Soc.Jpn.826_2013,Adachi_J.Phys.Soc.Jpn.8511_2016,Baqiya_Phys.Rev.B1006_2019}. 
As-grown samples were then annealed in vacuum under a pressure of \wave{} $10^{-6}$ Torr.
\Tc{} is defined as the crossing point of the zero-susceptibility line and the extrapolated line of the steepest part of the susceptibility curve.
Table~\ref{PLCCO_Conditions} summarizes key parameters of the annealing conditions and sample properties.
The sample reduction status, i.e., under-reduced (UR), optimally reduced (OP), and over-reduced (OR), is judged from 
\Red{
the annealing conditions, Ce content $x$, systematic changes of the c-axis length obtained by X-ray diffraction, and \Tc{} values.
As it is more difficult to remove excess oxygen with low Ce concentration, the same annealing conditions may result in under-reduced and over-reduced statuses for samples with low and high Ce content, respectively.}
The annealing consists of three steps, namely, protect annealing (PA)~\cite{Adachi_J.Phys.Soc.Jpn.826_2013,Horio_Nat.Commun.7_2016}, low-temperature annealing (LTA)~\cite{Krockenberger_Sci.Rep.3_2013}, and dynamic annealing (DA)~\cite{Wang_Phys.Rev.B809_2009, Sumura_Proc.14thInt.Conf.MuonSpinRotat.Relax.Reson.SR201721_2018, Ohnishi_J.Phys.Soc.Jpn.874_2018}. 
In the dynamic annealing, the annealing processes are separated into a few cycles providing sufficient time for oxygen atoms to diffuse from bulk to surfaces and thus further improves the oxygen homogeneity of the samples and increases the \Tc{}.
ARPES measurements were performed at beamline BL-28A of Photon Factory (PF) and BL5U of UVSOR facility with the total energy resolution of 30 meV. 
The samples were cleaved and measured \textit{in situ} at temperatures below 10 K and pressure better than \num{1.5d-10} Torr.
The photon energy was set at 55 eV with circular polarization at PF and at 60 eV with linear (perpendicular to the cut) polarization at UVSOR.

\begin{figure}[!bt]
    \includegraphics[width=0.48\textwidth]{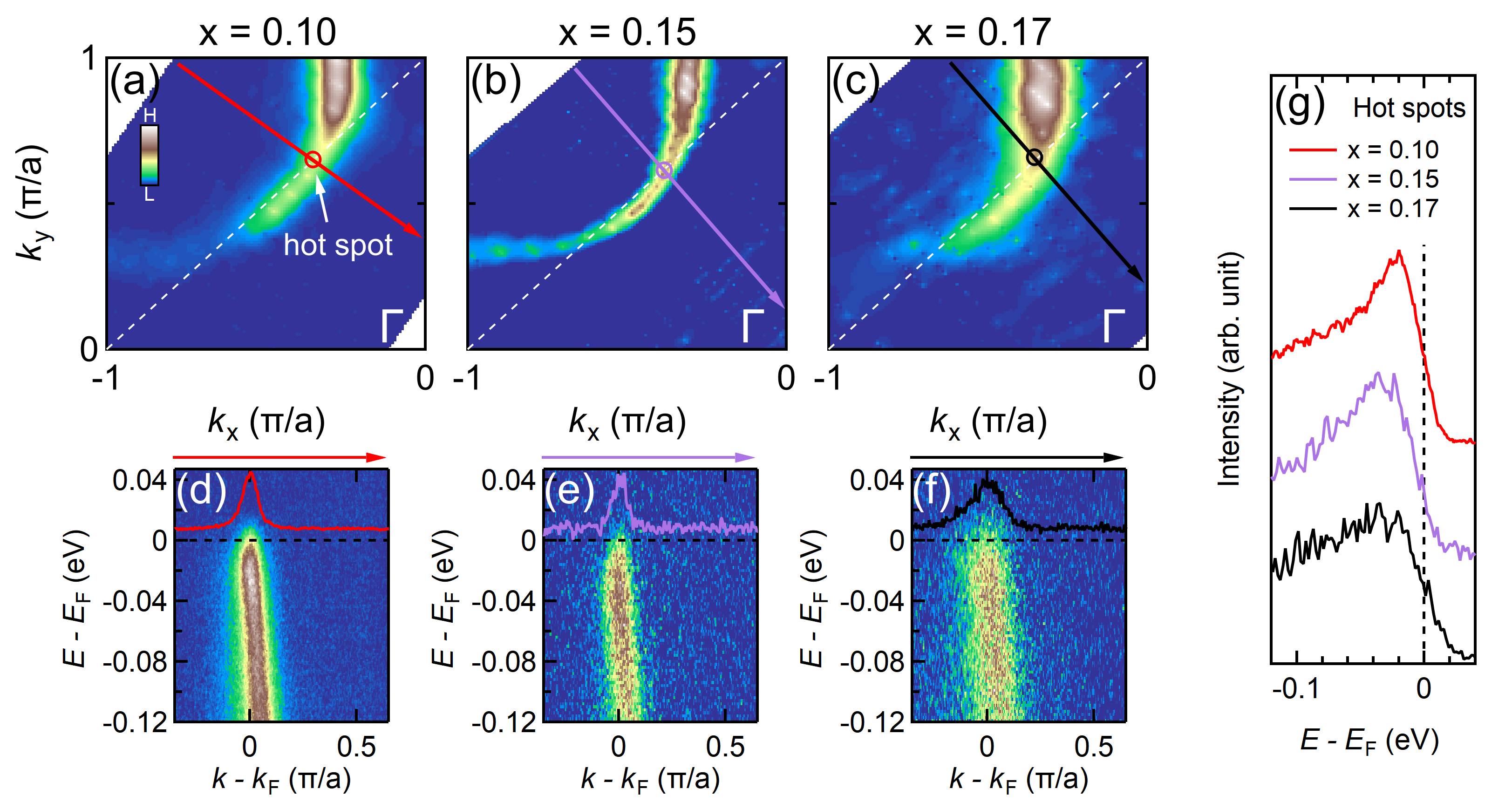}
    \caption[Doping dependence of band dispersions at the hot spots]{
    \Red{
    Doping dependence of band dispersion across the hot spots.
    (a) - (c) Magnified Fermi surfaces corresponding to Figs.~\ref{Fermi_Surface} (a) - \ref{Fermi_Surface} (c).
    The circles indicate the crossing points of the AFM Brillouin-zone boundaries (white dash lines) and the Fermi surfaces referred to as hot spots.
    (d) - (f) Band dispersions and MDCs at \EF{} along the momentum cuts indicated in (a) - (c), respectively.
    (g) Energy distribution curves (EDCs) at the hot spots in (a) - (c), respectively.
    }
    \label{Dispersions}
    }
\end{figure}

Figures~\refer{Fermi_Surface}{a} - \refer{Fermi_Surface}{c} show the ARPES Fermi-surface intensity plots of the protect-annealed PLCCO samples for $x$ = 0.10 with \Tc{} = 28 K, $x$ = 0.15 with \Tc{} = 19 K, and $x$ = 0.17 with \Tc{} = 6 K (La1.0), respectively.
\Red{
Figure~\refer{Fermi_Surface}{d} shows a representative experimental dispersion and momentum distribution curve (MDC) at \EF{} along the arrow indicated in Fig.~\refer{Fermi_Surface}{a} for PLCCO with $x$ = 0.10.
The red curves in Figs.~\refer{Fermi_Surface}{a} - \refer{Fermi_Surface}{d} are the tight-binding fitted bands for different doping levels, after following the same fitting procedure\textemdash taking advantage of the multi-peak Lorentzian fittings shown in Fig.~\refer{Fermi_Surface}{d}, we extract the experimental band dispersions, which are then fitted by the tight-binding model:
\begin{math}
\epsilon  = -2t(\cos k_xa+\cos k_ya) + 4t'\cos k_xa \cos k_ya - 2t''(\cos 2k_xa+\cos 2k_ya) + \epsilon_0,
\end{math}
where $\epsilon_0$ is the chemical potential, $t$, $t'$, and $t''$ are the transfer integrals between the nearest-neighbor, next-nearest-neighbor, and third-nearest-neighbor Cu atoms.
The fitted tight-binding parameters are listed in Table~\ref{TB_Parameters}.
Given the fitted coefficients, we plot the experimental band dispersions (red markers) and their tight-binding fitted dispersions (false-color contours) in Figs.~\refer{Fermi_Surface}{e} - \refer{Fermi_Surface}{g}.
Our previous study~\cite{Horio_Nat.Commun.7_2016} reveals that AFM correlation is strongly suppressed in the protect-annealed PLCCO single crystals, unlike those studies in conventionally annealed PLCCO (La1.0)~\cite{Song_Phys.Rev.Lett.11813_2017,Matsui_Phys.Rev.Lett.951_2005} or other electron-doped cuprates~\cite{Armitage_Phys.Rev.Lett.8714_2001,Matsui_Phys.Rev.B7522_2007,Ikeda_Phys.Rev.B801_2009}.
To confirm this, we present in Fig.~\ref{Dispersions} the doping dependence of dispersions across the hot spots.
One can identify that there is no signature of the AFM pseudogap at the hot spots for the doping range studied in the present work.
}
According to the Luttinger theorem~\cite{Luttinger_Phys.Rev.1194_1960}, the number of conduction electrons is proportional to the Fermi surface volume.
We then use the fitted tight-binding Fermi surfaces as summarized in Fig.~\refer{Fermi_Surface}{h} to calculate the Fermi-surface area 1 + \nFS{} and estimate the actual doped electron concentration \nFS{}~\cite{Wei_Phys.Rev.Lett.11714_2016,Horio_Nat.Commun.7_2016,Song_Phys.Rev.Lett.11813_2017}, as shown in Table~\ref{TB_Parameters}.
Error bars for \nFS{} are estimated by the propagation law of errors of the fitting coefficients~\cite{Horio_Nat.Commun.7_2016}.

\begin{table*}[!tb]
    \caption[Tight-binding parameters from global fitting]{
    \Red{
    Tight-binding parameters from the global fitting.
    Uncertainties are determined by the fitting errors.
    }
    \label{TB_Parameters}
    }
    \begin{ruledtabular}
    \begin{tabular}{ccccccc}
    {\textit{x}} &$T_{c}$ {(K)} &\textit{t} {(meV)} &\textit{t$^\prime$} {(meV)} &\textit{t$^{\prime\prime}$} {(meV)} &\textit{$\epsilon_0$} {(meV)} &\nFS{}\footnote{Error bars are estimated by the propagation law of errors of the fitting coefficients~\cite{Horio_Nat.Commun.7_2016}.
}\\
    \hline
    0.10 &26 &211 $\pm$ 3 &87 $\pm$ 6 &0 $\pm$ 3 &46 $\pm$ 10 &0.186 $\pm$ 0.016\\
    0.15\footnote{Parameters were averaged for four samples with the same Ce concentration.} &18 &193 $\pm$ 10 &63 $\pm$ 9 &10 $\pm$ 4 &-35 $\pm$ 14 &0.242 $\pm$ 0.024\\
    0.17 &6 &163 $\pm$ 5 &55 $\pm$ 8 &26 $\pm$ 3 &-58 $\pm$ 13 &0.250 $\pm$ 0.022\\
    \end{tabular}
    \end{ruledtabular}
\end{table*}

\begin{figure}[!bt]
    \includegraphics[width=0.48\textwidth]{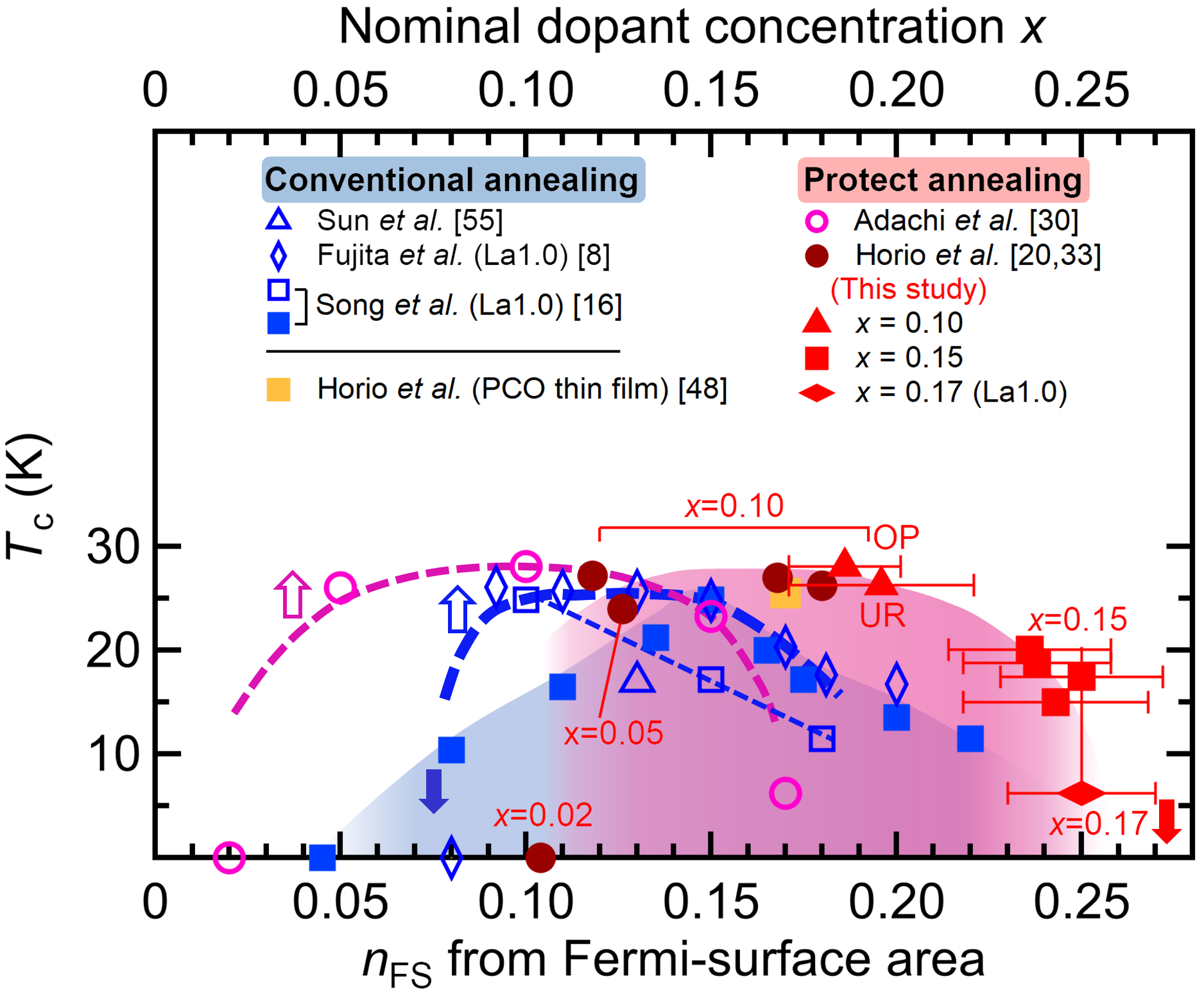}
    \caption[New electron-concentration-based phase diagram of protect-annealed PLCCO]{
    New electron-concentration (\nFS{})-based phase diagram of protect-annealed PLCCO.
    The \Tc{} values of PLCCO with $x$ = 0.10 [optimally-reduced (OP) and under-reduced (UR)], $x$ = 0.15 \Green{\Strike{[over-reduced (OR)]}}, and PLCCO (La1.0) with $x$ = 0.17 are plotted against \nFS{} [estimated from the Fermi-surface area shown in Fig.~\refer{Fermi_Surface}{h}] as filled red markers.
    Error bars are estimated by the propagation law of errors of fitting coefficients~\cite{Horio_Nat.Commun.7_2016}.
    Our previous results on protect-annealed PLCCO with $x$ = 0.02, 0.05, and 0.10 are also plotted (filled dark red circles, Horio \etal{}~\cite{Horio_Nat.Commun.7_2016,Horio_ArXiv180104247Cond-Mat_2018}), which, together with the current results, reveal a \nFS{}-based superconducting (SC) dome (magenta shaded area).
    For the $x$ = 0.17 data point, a vertical line is drawn from the \Tc{} defined as described in the text to the $T_\text{c,onset}$. 
    For all the other samples, the \Tc{} and $T_\text{c,onset}$ are close to each other and fall within the size of the markers.
    The \Tc{} values of OP samples are also plotted traditionally against Ce concentration (empty magenta circles, Adachi \etal{}~\cite{Adachi_Condens.Matter23_2017}).
    Data based on conventionally annealed PLCCO and PLCCO (La1.0) are included for comparison and indicated by blue markers (the empty markers are plotted against Ce concentration~\cite{Sun_Phys.Rev.Lett.924_2004,Fujita_Phys.Rev.B671_2003,Song_Phys.Rev.Lett.11813_2017} and the filled markers are plotted against \nFS{} forming a blue shaded SC dome~\cite{Song_Phys.Rev.Lett.11813_2017}).
    The estimated \nFS{} and \Tc{} for annealed \Tp{}-type ${\text{Pr}_\text{2}\text{CuO}_\text{4}}$ (PCO) thin films~\cite{Horio_Phys.Rev.B(RapidComm.)982_2018} is also plotted as an orange filled square.
    \Red{\Strike{The green curve indicates the SC dome of hole-doped T-type LSCO~\cite{Takagi_Phys.Rev.B404_1989} after scaling the maximum \Tc{} value to that of the protect-annealed PLCCO.}}
    Dashed curves and lines are the guide to the eyes.
    \label{New_Phase_Diagram}
    }
\end{figure}

In Fig.~\ref{New_Phase_Diagram}, by plotting the \Tc{} values against \nFS{} for multiple samples (filled red markers), we show the new \nFS{}-based phase diagram (the SC dome shown by a magenta shaded area) of the protect-annealed PLCCO with $x$ = 0.10, 0.15, and 0.17 (La1.0).
Our previous results on PLCCO samples with $x$ = 0.02 (non-SC), 0.05 (\Tc{} = 24 K), and 0.10 (\Tc{} = 27 K) are included as filled dark red circles~\cite{Horio_Nat.Commun.7_2016,Horio_ArXiv180104247Cond-Mat_2018}.
Meanwhile, the \Tc{} values of OP protect-annealed PLCCO are also plotted against Ce concentration (empty magenta circles)~\cite{Adachi_Condens.Matter23_2017}.
One can immediately see the difference between the traditional Ce-based phase diagram (magenta dashed curve) and the \nFS{}-based (magenta shaded area), i.e., \nFS{} estimated from the Fermi-surface area is significantly larger than the Ce doping level $x$.
The remarkable increase of actual electron concentration by protect annealing yields an extension of the SC dome on the overdoped side.
For comparison, the \Tc{} values of conventionally annealed PLCCO and ${\text{Pr}_{\text{1.0-}x}\text{La}_\text{1.0}\text{Ce}_x\text{CuO}_\text{4}}$ [denoted as PLCCO (La1.0)] samples are also plotted by blue markers, with empty markers plotted against the Ce concentration~\cite{Sun_Phys.Rev.Lett.924_2004,Fujita_Phys.Rev.B671_2003,Song_Phys.Rev.Lett.11813_2017} and filled markers plotted against \nFS{}~\cite{Song_Phys.Rev.Lett.11813_2017}.
The thick dashed blue curve tracks the data from Fujita \etal{}~\cite{Fujita_Phys.Rev.B671_2003} showing examples of the traditional phase diagram for electron-doped cuprates (Fig.~\ref{Phase_Diagram}).
Owing to the improved annealing method, superconductivity in the protect-annealed PLCCO samples can be realized not only with higher \Tc{} but also at a Ce doping level as low as 0.05, which has never been reported for the conventional annealing method.
We also replot the results from Song \etal{}, who obtained a \nFS{}-based SC dome (blue shaded area enclosed by filled blue squares) by conventionally annealing and oxidizing PLCCO (La1.0) samples with three Ce concentrations (empty blue squares tracked by a thin dashed blue line)~\cite{Song_Phys.Rev.Lett.11813_2017}.
Apparently, the SC dome obtained in protect-annealed PLCCO is more extended on the overdoped side than that based on the conventional annealing method.
Quantitatively, for protect annealed PLCCO and conventionally annealed PLCCO (La1.0), the addition of actual electron concentration (\nFS{} \textendash ~$x$) is \wave{} 0.08 and 0.04 $e$/Cu, respectively, in average.
This indicates the improved protect (and dynamic) annealing method dopes the system with more electrons than the conventional annealing method.
It is likely that the reduction of the excess apical oxygen together with those oxygen atoms at the regular sites (oxygen sites in the (La,Pr)O and CuO$_2$ layers) and the improvement of oxygen homogeneity give rise to the significant amount of additional electrons~\cite{Horio_Nat.Commun.7_2016,Adachi_Condens.Matter23_2017}.

Figure~\ref{New_Phase_Diagram} also includes the data on the annealed \Tp{}-type ${\text{Pr}_\text{2}\text{CuO}_\text{4}}$ (PCO) thin films~\cite{Horio_Phys.Rev.B(RapidComm.)982_2018} as an orange filled square.
It turns out that the data point of PCO thin films falls on the \nFS{}-based SC dome of the protect-annealed PLCCO, supporting the validity of the new phase diagram which employs \nFS{} as the doping axis.
Nevertheless, further studies on other electron-doped systems such as ${\text{Nd}_{\text{2-}x}\text{Ce}_x\text{CuO}_\text{4}}$ are needed to establish the new phase diagram.
We note that, as \Tc{} drops rapidly on the underdoped side of the \nFS{}-based SC dome, the AFM correlation is found to coexist with superconductivity in the $x$ = 0.05 samples~\cite{Horio_ArXiv180104247Cond-Mat_2018}.
At present, we are not sure whether this drop of \Tc{} and the coexistence of the strong AFM correlation and superconductivity is intrinsic or not.
Because of the dramatic suppression of the AFM phase in the protect-annealed PLCCO with $x$ = 0.10~\cite{Horio_Nat.Commun.7_2016}, one may anticipate a \Tc{} recovery with further improved annealing methods for $x \leq 0.05$.
\Red{Finally, if one compares the present results with those of the hole-doped counterpart, the SC dome of PLCCO on the overdoped side is as widely extended as that LSCO~\cite{Takagi_Phys.Rev.B404_1989}.}
\Red{\Strike{we also replot the SC dome of LSCO~\cite{Takagi_Phys.Rev.B404_1989} as a green curve in Fig.~\ref{New_Phase_Diagram}.
Here, \Tc{} values are scaled to the maximum \Tc{} value of the protect-annealed PLCCO and plotted against Sr concentration $x$.}}
One sees that the doping range of the \nFS{}-based SC dome of the protect-annealed PLCCO is similar to that of the hole-doped LSCO, which may provide a clue for understanding the symmetry/asymmetry of the phase diagram of cuprate superconductors.


In summary, we have performed ARPES measurements on protect-annealed PLCCO single crystals with Ce concentrations $x$ = 0.10, 0.15, and 0.17 (La1.0). 
The actual electron concentration \nFS{} estimated from the Fermi-surface area is significantly larger than the nominal Ce doping level by \wave{} 0.08 $e$/Cu.
Owing to the improved protect (and dynamic) annealing method, which dopes the system with electrons, the new \nFS{}-based SC dome of PLCCO is more extended on the overdoped side than that based on the conventional annealing method.
The reduction of the excess apical oxygen together with those oxygen atoms at the regular sites and the improvement of oxygen homogeneity may be able to account for the significant amount of additional electrons.
The present results suggest that employing \nFS{} as the doping axis is a useful way when investigating the temperature-doping phase diagram of electron-doped cuprates.
Furthermore, the similar \Red{overdoped end points of the SC dome} between the new \nFS{}-based phase diagram of the electron-doped cuprates and that of hole-doped LSCO may provide a clue for understanding the electron-hole symmetry/asymmetry of the cuprate phase diagram.

\begin{acknowledgments}
ARPES experiments were performed at KEK Photon Factory (proposal nos. 2015S2-003, 2016G096, 2018G049, and 2018S2-001) and UVSOR (proposal nos. 28-813 and 29-821).
A part work of sample preparations was conducted at the Advanced Characterization Nanotechnology Platform of the University of Tokyo, supported by "Nanotechnology Platform" of MEXT, Japan.
This work was supported by KAKENHI Grants (Nos. 14J09200, 15H02109, 17H02915, 19K03741, and 19H01841) from JSPS and "Program for Promoting Researches on the Supercomputer Fugaku" (Basic Science for Emergence and Functionality in Quantum Matter) from MEXT.
\end{acknowledgments}

\bibliography{PLCCO}

\end{document}